\documentclass[preprint,showpacs,preprintnumbers,amsmath,amssymb,superscriptaddress, nofootinbib]{revtex4}  
\usepackage{graphicx,color}
\usepackage{amsmath,amssymb}
\usepackage{url}
\usepackage{epstopdf}
\usepackage{slashed}

\newcommand{\be}{\begin{equation}}
\newcommand{\ee}{\end{equation}}
\newcommand{\bea}{\begin{eqnarray}}
\newcommand{\eea}{\end{eqnarray}}

\newcommand{\crn}{\nonumber \\}

\newcommand{\bc}{\begin{center}}
	\newcommand{\ec}{\end{center}}

\newcommand {\ba}{\begin{array}}
	\newcommand {\ea}{\end{array}}
\newcommand{\ben}{\begin{enumerate}}
	\newcommand{\een}{\end{enumerate}}
	
\usepackage{epsfig,graphicx} 
\usepackage{bm}
\usepackage{dcolumn}
\begin{document}
	\title{Decays of Standard Model like Higgs boson $h \rightarrow\gamma\gamma, Z \gamma$ in a minimal left-right symmetric model}
	\author{T.T. Hong }
\affiliation{An Giang University, Long Xuyen City, Vietnam} 

\affiliation{Vietnam National University, Ho Chi Minh City, Vietnam}

\author{V.K. Le}
\affiliation{An Giang University, Long Xuyen City, Vietnam} 
\affiliation{Binh Thuy Junior High School, Bui Huu Nghia Street, Binh Thuy Ward, Binh Thuy District, Can Tho City, Vietnam}

\author{L.T.T. Phuong }
\affiliation{An Giang University, Long Xuyen City, Vietnam}

\author{N.C.Hoi }
\affiliation{An Giang University, Long Xuyen City, Vietnam}

\author{N.T.K. Ngan }
\affiliation{Department of Physics, Can Tho University,
	3/2 Street, Can Tho, Vietnam}

\author{N.H.T. Nha}
\email{nguyenhuathanhnha@vlu.edu.vn}
\affiliation{Subatomic Physics Research Group, Science and Technology Advanced Institute, Van Lang University, Ho Chi Minh City, Vietnam
}
\affiliation{Faculty of Applied Technology, School of Engineering and Technology,  Van Lang University, Ho Chi Minh City, Vietnam}

\begin{abstract}
Two decay channels $h\rightarrow \gamma \gamma, Z\gamma$ of the Standard Model-like Higgs in a left-right symmetry model  are investigated under recent experimental data. We will show there exist one-loop contributions that affect the $h\rightarrow Z\gamma$ amplitude, but not the $h\rightarrow \gamma\gamma$ amplitude. From numerical investigations, we show that the signal strength $\mu_{Z \gamma}$ of the decay  $h\rightarrow Z\gamma$ is still constrained strictly by that of $h\rightarrow \gamma\gamma$, namely $|\Delta \mu_{\gamma \gamma}|<38\%$ results in max $|\Delta \mu_{Z \gamma}|<46\%$.  On the other hand, the future experimental sensitivity $|\Delta \mu_{\gamma \gamma}|=4\%$ still allows $|\Delta \mu_{Z \gamma}|$ reaches to values  larger than the expected sensitivity $|\Delta \mu_{Z \gamma}|=23\%$.
\end{abstract}

\maketitle

\section{\label{intro} Introduction}
\allowdisplaybreaks
 The  standard model-like (SM-like) Higgs decay $h\to Z \gamma$   is one of the most important channels being searched at the experimental center  \cite{CMS:2022dwd}. Meanwhile, the experimental evidence of  this loop-induced  decay relating to the effective coupling $hZ\gamma$  has been reported by ATLAS and CMS recently \cite{CMS:2022ahq, CMS:2023mku}, in agreement with the SM prediction within 1.9 standard deviation. Experimental data shows  that the effective coupling $h\gamma \gamma$ derived from $h\to \gamma\gamma$ decay rates is   constrained very  strictly  \cite{CMS:2021kom}. In contrast, the  effective coupling $hZ\gamma$ in many models beyond SM (BSM) might  differ considerably from the SM prediction, because the $Z$ couplings  to new particles are less strict than those of the photon. Hence, studying the effective $hZ\gamma$ couplings will be an indirect channel to determine the properties of new particles. Controlled by the strict experimental constraint of the decay $h\to \gamma \gamma$, constraints of the SM-like Higgs decay  $h\to Z \gamma$ affected by new fermions and charged scalars were studied in several BSMs such as 3-3-1 models \cite{Yue:2013qba, Hung:2019jue}, only Higgs extended SM versions \cite{Fontes:2014xva, Yildirim:2021kqs, Benbrik:2022bol, Hue:2023tdz}, $U(1)$ gauge   extensions from SM \cite{Wang:2022fug, Tran:2023vgk}, supersymmetric models \cite{Archer-Smith:2020gib, Liu:2020nsm, Cao:2013ur}, chiral extension of the SM \cite{Barducci:2023zml}, $\dots$.  Previous studies $h\to Z \gamma$  in left-right symmetric models  ignored one-loop contributions relating to the diagrams consisting of  both virtual Higgs and gauge particles in the loops \cite{Martinez:1990ye, Martinez:1989kr}, where the $h$-Higgs-gauge boson couplings were assumed to be suppressed.    

The  experimental results have been updated for loop-induced Higgs  decays   $h\rightarrow\gamma\gamma$ \cite{Aaboud:2018ezd, Sirunyan:2018ouh, Aaboud:2018xdt}  and $h\rightarrow Z\gamma$ \cite{Aaboud:2017uhw}. In the future project, the significant strength of the decay $h\rightarrow Z\gamma$, denoted as $\mu_{Z\gamma}$, can reach $\Delta \mu_{Z\gamma} \equiv\mu_{Z\gamma}-1= \pm0.23$, while that of the channel $h\rightarrow \gamma\gamma$ can reach around $\Delta \mu_{\gamma \gamma}\equiv \mu_{\gamma \gamma}-1=\pm 0.04$ from two  CMS  and ATLAS experiments \cite{Cepeda:2019klc}.  In addition, the ATLAS expected significance at HL-LHC  to the $h\rightarrow\, Z\gamma $ channel will  be $4.9~\sigma$ with $3000~\mathrm{fb}^{-1}$.  Also,   the Circular Electron Positron Collider (CEPC) \cite{An:2018dwb} can reach a sensitivity of  $ \mu_{Z\gamma} =1\pm0.22$ \cite{Antonov:2022zwm}.

One interesting extension  of the beyond the SM models is  an extension of  the lepton sector. Namely, the minimal left-right version (MLRSM) is  constructed based on the parity symmetry $ SU{(2)_L}  \otimes SU(2)_R \otimes U(1)_{B-L}$  \cite{Pati:1974yy,Mohapatra:1974gc,Senjanovic:1975rk}, which contains  Higgs fields included in  two  $SU(2)_L$   triplets denoted as  $\Delta_{L,R}$ and a bi-doublet field $\Phi$ playing the SM  Higgs role. Therefore, the MLRSM allows us to  solve  the parity problem of the SM as well as   the neutrino oscillation data  through the seesaw mechanism. Besides, it contains extended particles which may result in interesting consequences for rare decays such as Higgs boson decay $h \to Z \gamma$.

This work is organized as follows. In section \ref{sec:model}, we present the overview of the MLRSM,  including  the particle content and physical states.  In section \ref{sec:coupling}, we present the necessary couplings that generate  one-loop contributions to the decays  $h\rightarrow \gamma \gamma, Z \gamma$. We will also collect analytic formulas to determine the decay rates, reminding some new contributions that were not discussed previously. Numerical results will be  investigated  in section \ref{sec:numerical}. Namely, we will investigate the dependence of $\mu_{Z \gamma}^{\mathrm{MLRSM}}$ on several important parameters in this model. Finally, the summary is given in section \ref{sec:conclusion}.

\section{\label{sec:model}The minimal {left-right} symmetry model }

\subsection{The model review}
All  needed ingredients relevant to one-loop contributions to the decay amplitudes  $h \to Z \gamma,\gamma \gamma$ will be collected in this section. In most general, the electric charge operator can be written as \cite{Zhang:2007da, Lee:2017mfg}
\bea
Q=T^R_3 +T^L_3+\dfrac{B-L}{2}=T^{L,R}_3+\dfrac{Y}{2},
\label{eq:charge_Q}
\eea
where  $T^{L,R}_i$ are the generators of the gauge groups $SU(2)_{L,R}$; $B$ ($L$)  is   the baryon (lepton) number  defining the $U(1)_{B-L}$ group in the MLRSM.  
The baryon and lepton number of the fermions can be written in the table \ref{t_BL}.
\begin{table}[h]
	\centering
	\begin{tabular}{|c|c|c|}
\hline
$f$&$e,\mu,\tau,\nu_e,\nu_\mu,\nu_\tau$&$u,c,t,s$ \\
\hline
$L$&1&0 \\
\hline
$B$&0&$\dfrac{1}{3}$ \\
\hline
\end{tabular}
\caption{The baryon and the lepton numbers of the fermions in MLRSM \label{t_BL}}
\end{table}

With this information, we can write down the lepton and fermion representations as follows
\begin{equation}	
{L'_{Li}}=\begin{pmatrix}
\nu'_{Li}\\
l'_{Li}\\
\end{pmatrix}\thicksim (2,1,-1),\
{L'_{Ri}}=\begin{pmatrix}
\nu'_{Ri}\\
l'_{Ri}\\
\end{pmatrix}\thicksim (1,2,-1),
\end{equation}
\begin{equation}	
{Q'_{Li}}=\begin{pmatrix}
u'_{Li}\\
d'_{Li}\\
\end{pmatrix}\thicksim (2,1,\dfrac{1}{3}),\
{Q'_{Ri}}=\begin{pmatrix}
u'_{Ri}\\
d'_{Ri}\\
\end{pmatrix}\thicksim (1,2,\dfrac{1}{3}),
\end{equation}\\
where $i=1,2,3$ is the flavor index. 

 Gauge boson and fermion masses  are originated from  the following scalar  sector, consisting of a bi-doublet  and two triplet scalar fields $\Delta_{L,R}$  satisfying
\begin{equation}	
\Phi=\begin{pmatrix}
\phi^{0}_{1}&\phi^{+}_{2}\\
\phi^{-}_{1}&\phi^{0}_{2}\\
\end{pmatrix}, ~ \widetilde{\Phi}=\sigma_2\Phi^*\sigma_2 \thicksim (2,2,0),\; \Delta_{L,R}=\begin{pmatrix}
\dfrac{\delta^{+}_{L,R}}{\sqrt{2}}&\delta^{++}_{L,R}\\
\delta^{0}_{L,R}&-\dfrac{\delta^{+}_{L,R}}{\sqrt{2}}\\
\end{pmatrix}\thicksim (3,1,2).
\end{equation}

The Higgs components develop vacuum expectation values (VEV) defined as 
\begin{equation}	
\langle \Phi \rangle=\begin{pmatrix}
\langle \phi^0_1\rangle&0\\
0&\langle \phi^0_2\rangle\\
\end{pmatrix};\
\langle \Delta_{L,R} \rangle=\begin{pmatrix}
0&0\\
\langle \delta^0_{L,R} \rangle &0\\
\end{pmatrix},
\end{equation}
where the neutral Higgs components are expanded as follows 
\begin{align}\label{vevhigg1}  
\phi^0_i&= \langle \phi^0_i\rangle +\frac{ r_i +i a_i}{\sqrt{2}} , \langle \phi^0_1\rangle= \frac{k_1}{\sqrt{2}},   \; \langle \phi^0_2\rangle= \frac{k_2 e^{i\alpha}}{\sqrt{2}} , \; i=1,2;
\crn   \delta^0_{L,R} &=\langle \delta^0_{L,R} \rangle + \frac{v_{L,R}e^{i\theta_{L,R}}+ r_{L,R} +i a_{L,R}}{\sqrt{2}},\; \langle \delta^0_{L} \rangle =\frac{v_Le^{i\theta_L}}{\sqrt{2}}, \; \langle \delta^0_{R} \rangle =\frac{v_R}{\sqrt{2}}. 
\end{align}

The symmetry breaking pattern in MLRSM happens in two following steps:  
$SU(2)_L\otimes SU(2)_R \otimes  U(1)_{B-L}\xrightarrow{v_R \ne 0, k_1, k_2, v_L=0} SU(2)_L\otimes U(1)_Y\xrightarrow{k_1, k_2, v_L \ne 0} U(1)_Q$, which corresponds to the reasonable limits  that $v_R\gg k_1,k_2 \gg v_L$.  Only new gauge bosons will be massive after the first step. The second step is the SM symmetry-breaking generating masses for  the SM particles. When the symmetry is broken to step two, only $U(1)_Q$ remains unbroken, where $Q$ is the quantifier. As a result, the photon $A_\mu$ has no mass. We stress that the MLRSM contains  no more than three scalar multiplets ($\phi, \Delta_{L,R}$).  The physical spectrum and masses of all particles in the model under consideration are summarized as follows. 
\subsection{Fermions}
  Physical fermion states and their masses   always relate to the Yukawa interactions, which are included in the following Lagrangian parts for leptons and quarks:   
\begin{align}
\mathcal{L}_{\ell}^Y&=-\overline {L'_{Li}}\left( f^e_{ij}\Phi +\widetilde{f}^e_{ij}\widetilde{\Phi}\right) L'_{Rj} - \sum_{X=L,R}Y^e_{X,ij} \overline {L^{'c}_{Xi}}i\sigma_2\Delta_X L'_{Xj} +\text{h.c}, \crn 	
\mathcal{L}^Y_{q}&= -\overline{Q'_{Li}}\left( f^q_{ij}\Phi +\widetilde f^q_{ij}\widetilde{\Phi}\right) Q'_{Rj} +\text{h.c.}. \label{quark}
\end{align}

Then, the mass terms for leptons and quarks are computed. We will use the results for fermion masses and mixing presented in Refs. \cite{Zhang:2007da,Lee:2017mfg}, i.e. all the original and the physical  states of fermions are the same. They are identified with the SM ones and will be denoted as $e_{aL,R},u_{aL,R}$ and $d_{aL,R}$ in this work. The  mass matrices $M_{\ell}$ and $M_{u,d}$ for charged leptons and up and down quarks are 
\begin{align}
\label{eq:Mud}
M_{\ell}&= \frac{k_1 f^e +k_2 \tilde{f}^e}{\sqrt{2}}=  \frac{ k\left( c_{\beta} f^e + s_{\beta} \tilde{f}^e \right)}{\sqrt{2}}, 
\crn M_{u}&= \frac{k_1 f^q+k_2 \tilde{f}^q}{\sqrt{2}}= \frac{k \left(s_\beta  f^q  +c_{\beta}\tilde{f}^q\right) }{\sqrt{2}}, 
\crn \; M_{d}&= \frac{k_1 f^q+k_2 \tilde{f}^q}{\sqrt{2}}=  \frac{ k\left( c_{\beta} f^q + s_{\beta} \tilde{f}^q\right)}{\sqrt{2}},
\end{align}
where 
\begin{align}
	\label{eq:defk}
k^2\equiv k_1^2 +k_2^2,\; t_{\beta} \equiv \frac{s_{\beta}}{c_{\beta}}=\frac{k_1}{k_2}. 
\end{align}
As we will show below, the matching condition to the SM leads to $k=246$ GeV and the Yukawa couplings of quarks defined in the SM can be seen as follows $c_{\beta} f^e + s_{\beta} \tilde{f}^e  \to y^e$, $s_\beta  f^q  +c_{\beta}\tilde{f}^q\to y^u$ and  $ k_1 f^q+k_2 \tilde{f}^q \to y^d$. Here, we fix $\alpha=0$ so that the value of $t_{\beta}=s_{\beta}/c_{\beta}$ can be small, namely $t_{\beta} \ge 1.2$  \cite{Dekens:2021bro}. The three above fermion mass matrices are denoted as $M_f$ with $f=\ell,u,d$ can be diagonalized by two unitary transformations $V_f^{L}$ and $V_f^{R}$ as follows: $V^{L \dagger}_f M_f V^{R}_f= \hat{M}_f =\mathrm{diag}\left( m_{f,1},\; m_{f,2}, \;m_{f,3}\right)$. Here $m_{f,i}$ with $i=1,2,3$  and $f=\ell,u,d$ denotes the physical masses of charged leptons, up and down quarks.  The transformations between the flavor basis $f'_{L(R)}=(f'_{1},\; f'_{2},\; f'_{3})^T_{L(R)}$ and the mass basis $f_{L(R)}=(f_{1},\; f_{2},\; f_{3})^T_{L(R)}$ is $f'_{L(R)} =V_f^{L(R)} f_{L(R)}$.  As we will show below, the couplings of the SM-like Higgs boson with charged leptons and quarks are the same as the SM results.

\subsection{Gauge bosons}
   The  covariant derivative corresponding to the symmetry of the MLRSM is defined as   \cite{Zhang:2007da} 
\begin{equation}
D_\mu=\partial_\mu -ig_L\sum_{i=1}^{3}T^a_L W^a_{L\nu} +ig_R\sum_{i=1}^{3}T^a_R W^a_{R\nu} -ig'\dfrac{B-L}{2}B_\mu,
\end{equation}	
where $g_{L,R}$ and $g'$ are the  $SU(2)_{L,R}$ and $U(1)_{B-L}$ gauge couplings, respectively. 

The Lagrangian for scalar kinetic parts  is written as
\begin{align}\label{eq_LkH}
\mathcal{L}_s =&	 \sum_{S=\Phi,\Delta_{L,R}}\mathcal{L}_S \crn
=& \sum_{S=\Phi,\Delta_{L,R}}\mathrm{Tr}\left[ \left( D_\mu S\right) ^\dagger\left( D^\mu S\right) \right].
\end{align}	
The particular forms of covariant derivatives to the scalar multiplets are
\begin{align}
D_\mu\Phi&=\partial_\mu\Phi -ig_L\dfrac{\sigma^b}{2}W^b_{L\nu}\Phi+ig_R\Phi\dfrac{\sigma^a}{2}W^a_R,
\crn D_\mu \Delta_{X} &=  \partial_\mu \Delta_{X} -i\frac{g_{X}}{2}\left( \sigma^a\Delta_{X} - \Delta_{X}  \sigma^a\right) W^a_{X}-ig'I_2 B_\mu\Delta_{X}, 
\end{align}
where $X=L,R$, $\sigma^a$ is the Pauli matrix corresponding to the $SU(2)$ doublet  representation of $T_{L,R}^a$ with $a=1,2,3$. Therefore, the mass terms of gauge bosons are derived from the vev of Higgs components as follows 
\begin{align}
\langle\mathcal{L}_s\rangle= &  \sum_{S=\Phi,\Delta_{L,R}} \langle\mathcal{L}_S\rangle \crn
=&\dfrac{1}{8}\left(k^2+4v^2_L\right)g^2_L W^{3\mu}_L W^3_{L\mu} - \dfrac{1}{4}g_Lg_Rk^2 W^{3\mu}_L W^3_{R\mu} \crn
&+ \dfrac{1}{8}\left(k^2 +4v^2_R\right)g^2_R W^{3\mu}_R W^3_{R\mu}-g_Lg'W^{3\mu}_LB^\mu - g_Rg' v^2_RW^{3\mu}_RB_\mu \crn
& +\dfrac{1}{2}g^{'2}(v^2_L+v^2_R)B^\mu B_\mu +\dfrac{1}{4}g^2_L\left(k^2+2v^2_L\right)W^{+\mu}_L W^{-}_{L\mu} - \dfrac{1}{2}g_Lg_Rk_1k_2 e^{i\alpha} W^{+\mu}_L W^{-}_{R\mu}\crn
&- \dfrac{1}{2}g_Lg_Rk_1k_2 e^{-i\alpha} W^{+\mu}_L W^{-}_{R\mu} +\dfrac{1}{4}g^2_R\left(k^2 +2v^2_R\right) W^{+\mu}_R W^{-}_{R\mu},
\end{align}
where $k$ is defined in Eq. \eqref{eq:defk}.
The  mass terms  of the neutral and charged gauge bosons read: 
\begin{align}
\mathcal{L}^{mass}_g&=\dfrac{1}{2}(W^{3\mu}_L W^{3\mu}_R B^\mu)\begin{pmatrix}
\dfrac{1}{4}\left(k_1^2+ k_2^2+4v^2_L\right)g^2_L& 
-\dfrac{1}{4}g_Lg_R(k^2_1+ k^2_2) & - g_Rg' v^2_L\\
-\dfrac{1}{4}g_Lg_R(k^2_1+ k^2_2)& \dfrac{1}{4}\left(k_1^2+ k_2^2+4v^2_R\right)g^2_R & - g_Rg' v^2_R\\
- g_Rg' v^2_L& - g_Rg' v^2_R & g^{'2}(v^2_L+v^2_R)\\
\end{pmatrix}
\begin{pmatrix}
W^3_{L\mu}\\
W^3_{R\mu}\\
B_\mu\\
\end{pmatrix}\crn
&+ (W^{+\mu}_L  W^{+\mu}_R) \begin{pmatrix}
\dfrac{1}{4}\left(k_1^2+ k_2^2+2v^2_L\right)g^2_L & -\dfrac{1}{2}g_Lg_Rk_1k_2 e^{i\alpha}\\
-\dfrac{1}{2}g_Lg_Rk_1k_2 e^{-i\alpha}&\dfrac{1}{4}\left(k_1^2+ k_2^2+2v^2_R\right)g^2_R\\
\end{pmatrix}
\begin{pmatrix}
W^{-}_{L\mu}\\
W^{-}_{R\mu}\\
\end{pmatrix},
\end{align}
 where 
$
W^{\pm}_{X \mu} \equiv  \dfrac{1}{\sqrt{2}}\left( W^1_{X \mu}  \mp iW^2_{X \mu} \right)$ with $X=L,R$.  The mixing angle $\xi $ between two singly charged gauge bosons $W^\pm_L$ and $W^\pm_R$  is determined  by the following formula  
\begin{align}
\tan2\xi   =\dfrac{-4 g_Lg_Rk_1k_2}{(g^2_R -g^2_L ) (k_1^2+ k_2^2)+ 2(g^2_R v^2_R - g^2_L v^2_L )}. 
\end{align}

 Using
the approximation that $\tan2\xi\ll 1 \Rightarrow \tan2\xi\approx \sin2\xi\approx  2\sin\xi \approx 2\xi$, and $v_L\ll k_1, k_2 \ll  v_R$,  the $W_L-W_R$ mixing angle  $\xi$  is    
\begin{align}
2\xi &=\dfrac{-2g_Lk_1k_2}{g_Rv^2_R},\; 
\dfrac{k^2}{v^2_R} \approx x \ll 1,\ \dfrac{v^2_L}{v^2_R}\approx 0,\ \dfrac{v^2_L}{k^2}\approx x_L\ll 1.
\end{align}

The singly charged gauge bosons $W_{L,R}^\pm$  can be written as functions of the mass basis  ($W_1^\pm, W_2^\pm$) as follows
\begin{align}
\begin{pmatrix}
W^{\pm}_{L}\\
W^{\pm}_{R}\\
\end{pmatrix}= 
\begin{pmatrix}
c_\xi&-s_\xi e^{i\alpha}\\
s_\xi e^{-i\alpha}&c_\xi\\
\end{pmatrix}
\begin{pmatrix}
W^{\pm}_1\\
W^{\pm}_2\\
\end{pmatrix},
\end{align}
where $c_\xi\equiv \cos\xi$ and $s_\xi\equiv \sin\xi$. The respective   charged gauge boson masses are found to be
\begin{align}	
& m^2_{W_1} \approx \dfrac{1}{4}k^2g^2_L , \quad  m^2_{W_2} \approx \dfrac{1}{2}g^2_Rv^2_R. 
\end{align}
Identifying the $W^\pm_1\equiv W^\pm$ in the SM, we get $k_1^2+k_2^2=k^2\equiv v^2=(246\; \mathrm{GeV})^2$. 

The original  neutral gauge  basis $(W^3_{L\mu}, W^3_{R\mu}, B_\mu)$  are expressed in terms of the mass basis $(A_{\mu}, Z_{1\mu},Z_{2 \mu})$ as follows 
\begin{align} \label{CT}
\begin {pmatrix}
W_{L\mu}^3\\
W_{R\mu}^3\\
B_\mu\\
\end{pmatrix}=C^T\begin{pmatrix}
A_\mu\\
Z_{1\mu}\\
Z_{2\mu}\\
\end{pmatrix},
\end{align}
where 
\begin{align}
&C=\begin{pmatrix}
s_{z_2}&c_R c_{z_2}& c_{z_2} s_R\\
c_{z_2} c_{z_3}& -c_R c_{z_3} s_{z_2} - s_R s_{z_3}& -c_{z_3} s_R s_{z_2} + c_R s_{z_3}\\
-c_{z_2} s_{z_3}&-c_{z_3} s_R + c_R s_{z_2} s_{z_3}& c_R c_{z_3} + s_R s_{z_2} s_{z_3}\\
\end{pmatrix}\crn
\end{align}
and the mixing {angles} $t_R, t_{z_2}, t_{{z}_3}$ are given by
\begin{align}
t_R = \frac{g_R}{g'},\quad t_{z_2} = \frac{g_Rg'}{g_L\sqrt{g'^2+g_R^2}},\quad  t_{{2z}_3}=\dfrac{-g_R^2  k^2 \sqrt{g_L^2 g^{'2} + g_L^2 g_R^2 + g^{'2} g_R^2}}{2v_R^2(g_R^2+g^{'2})}.
\label{matching_coupl} 
\end{align}

State synchronization with the SM as follows: $W^{3}_{L\mu}\equiv W^3_{\mu}$, $Z_{1\mu}\equiv Z_{\mu}$ in the limits $c_{z_3}\to 1$, then we also have 
\begin{align}
\label{eq_tz2}
s_{z_2}\equiv s_W,\; c_{z_2}\equiv c_W \Rightarrow t_W=t_{z_2}=s_R t'.
\end{align} 
The Weinberg angle $\theta_W$  is identified from the definition   $\cos \theta_W \equiv c_{W}\equiv\frac{m_{W_1}}{m_{Z_1}} \thickapprox c_{z_2} $. Then, the  neutral gauge boson masses of  $Z_1,Z_2,$ and the photon $ A$ are given by 
\begin{align}
m^2_{Z_1}&\thickapprox  \dfrac{g^2_L v^2}{4c^2_{W}}, ~m^2_{Z_2}\thickapprox \dfrac{g^2_R v^2_R}{1-(g^2_L/g^2_R)t^2_{W}}, \text{and}~ m_A^2=0. 
\end{align}
In addition,  $Z_1\equiv Z$, and $Z_2\equiv Z'$ are respectively  SM gauge boson $Z$  found experimentally, and the heavy one appearing in the MLRSM.

\subsection{Higgs bosons}
The MLRSM scalar potential  is written as \cite{Zhang:2007da} 
\begin{align} \label{Vh1}
V_h&= -\mu^2_\Phi \mathrm{Tr}[\Phi^{\dagger}\Phi] -\mu'^2_\Phi (\mathrm{Tr}[\Phi^{\dagger} \widetilde \Phi]+\mathrm{Tr}[\widetilde\Phi^{\dagger}\Phi] ) -\mu^2_\Delta \sum_{X=L,R} \mathrm{Tr}[\Delta^{\dagger}_X \Delta_X]\crn
&{+\lambda_1 (\mathrm{Tr}[\Phi^{\dagger} \Phi])^2+ \lambda_2 \Big((\mathrm{Tr}[\Phi^{\dagger}\widetilde\Phi])^2 +(\mathrm{Tr}[\widetilde\Phi^{\dagger}\Phi])^2\Big)}+\lambda_3(\mathrm{Tr}[\Phi^{\dagger}\widetilde\Phi]\mathrm{Tr}[\widetilde\Phi^{\dagger}\Phi] )\crn &+\lambda_4\mathrm{Tr}[\Phi^{\dagger}\Phi](\mathrm{Tr}[\Phi^{\dagger}\widetilde\Phi]+\mathrm{Tr}[\widetilde\Phi^{\dagger}\Phi])+\rho_1(\mathrm{Tr}[\Delta^{\dagger}_L \Delta_L]^2+\mathrm{Tr}[\Delta^{\dagger}_R \Delta_R]^2)\crn
& +\rho_2 \sum_{X=L,R}(\mathrm{Tr}[\Delta^{\dagger}_X \Delta^{\dagger}_X]{\mathrm{Tr}[\Delta_X\Delta_X]}) +\rho_3(\mathrm{Tr}[\Delta^{\dagger}_L \Delta_L]\mathrm{Tr}[\Delta^{\dagger}_R \Delta_R]) \crn
& + \sum_{X\neq Y=L,R}\rho_4(\mathrm{Tr}[\Delta^{\dagger}_X \Delta^{\dagger}_X]\mathrm{Tr}[\Delta_Y \Delta_Y] )
+ \alpha_1 \mathrm{Tr}[\Phi^{\dagger}\Phi] \sum_{X=L,R}(\mathrm{Tr}[\Delta^{\dagger}_X \Delta_X])\crn  
& +\left\lbrace\alpha_2e^{i\delta_2}\sum_{X=L,R}(\mathrm{Tr}[\Phi^{\dagger} \widetilde \Phi]\mathrm{Tr}[\Delta^{\dagger}_X \Delta_X] )+ \mathrm{h.c.}\right\rbrace  + \sum_{X=L,R} \left( \alpha_3 \mathrm{Tr}[\Phi\Phi^{\dagger}\Delta_X \Delta^{\dagger}_X]   \right) \crn
&+ \left\{ \alpha_4 \mathrm{Tr}[\Phi^{\dagger}\Delta^{\dagger}_L\Phi\Delta_R] + \alpha_5\mathrm{Tr}[\Phi^{\dagger}\Delta^{\dagger}_L \widetilde\Phi\Delta_R ]+ {\alpha_6} \mathrm{Tr}[ \widetilde\Phi^{\dagger}\Delta^{\dagger}_L\Phi\Delta_R] +\mathrm{h.c.} \right\}.
\end{align}

From the minimal conditions of the  Higgs potential given in Eq. \eqref{Vh1}, three parameters   $\mu_\Phi^2$, $\mu'^2_\Phi$, and $\mu_\Delta^2$ are expressed as functions of other independent parameters. Inserting them into the Higgs potential \eqref{Vh1}, we can determine all  Higgs boson masses and  physical states.  Firstly, the original and the mass base of neutral CP-even Higgs bosons are related to each other as follows 
\begin{align} \label{ptrh0H01}
\begin{pmatrix}
h^0 \\
H^0_{1}\\
\end{pmatrix}=\left(
\begin{array}{cc}
s_{\beta } & c_{\beta } \\
-c_{\beta } & s_{\beta } \\
\end{array}
\right) \begin{pmatrix}
r_1\\
r_2\\
\end{pmatrix}.
\end{align}
We not that Eq. \eqref{ptrh0H01} do not use the the limit $k_1\gg k_2$ mentioned in Ref. \cite{Zhang:2007da},  which gives $t_\beta=\dfrac{k_1}{k_2}\approx \dfrac{1}{\epsilon_2}\gg 1$, $c_\beta = \dfrac{1}{\sqrt{t^2_\beta +1} }\approx \dfrac{1}{t_\beta}\approx \epsilon_2$, $s_\beta\approx 1$, and $\epsilon_1\equiv k_1/v_R$, $\epsilon_2\equiv k_2/k_1$. Besides that from Eq.  \eqref{ptrh0H01} we get the same result as in
Ref.~\cite{Zhang:2007da} in this limit.  In this study, the  SM-like Higgs mass is calculated approximately  to the order $\epsilon^2=v^2/v_R^2$, namely 
\begin{align} \label{eq_m2h0ep2}
m^2_{h^0}=& \left[2
\lambda_1+ 8 c_{\beta }^2  s_{\beta }^2\left( 2\lambda_2 +  \lambda_3\right)   + 8c_{\beta }  s_{\beta } \lambda_4 \right] v^2 \crn 
& -\frac{8 \left(2 c_{\beta }^2-1\right)^3 v^4 }{\alpha_3 v_R^2} \left[ 4 c_{\beta }^4 (2 \lambda_2 +\lambda_3)^2 -4s_\beta c_{\beta } \lambda_4 (2 \lambda_2 +\lambda_3) -4 c_{\beta }^2 (2 \lambda_2+\lambda_3)^2 -\lambda_4^2 \right]. 
\end{align}

The SM-like Higgs property appears in Eq. \eqref{eq_m2h0ep2} as $m^2_{h^0}\simeq v^2\times \mathcal{O}(\frac{v^2}{v_R^2})\sim  v^2$ because $\mathcal{O}(\epsilon^2) \simeq0$ when $v^2\ll v_R^2$.  In this limit, $h^0\equiv\, h$ can be identified with the SM-like Higgs boson with mass $m_{h}= 125.38$ GeV confirmed experimentally  \cite{CMS:2022dwd}. Then the Higgs self-coupling  $\lambda_{1}$ is expressed as follows 
\begin{align} \label{eq_fla1}
\lambda_1 &= \frac{1}{2} \left( \frac{m^2_{h}}{v^2}-8 c_{\beta } \left[ s_{\beta}^2 c_{\beta }  (2 \lambda_2+\lambda_3)+\lambda_4 s_{\beta } \right] 
\right. \crn& \left. \qquad \quad -\frac{8 v^2 \left(2
	c_{\beta }^2-1\right)^3  \left[ 2 c_{\beta } s_{\beta } (2 \lambda_2+\lambda_3) +\lambda_4 \right]^2}{\alpha_3 v_R^2}\right). 
\end{align}
We note that Eq. \eqref{eq_m2h0ep2} for SM-like Higgs mass is consistent with Ref. \cite{Duka:1999uc, Chakrabortty:2016wkl, Mitra:2016kov},  implying  that the $m_{h}$ value  is still at the electroweak scale even in the case of large  $t_{\beta}$.  Therefore, the value of $t_{\beta} \geq 1.2$ is still allowed to get the SM-like Higgs mass consistent with the experiment.

Regarding the SM-like Higgs couplings with charged leptons and fermions,  using Eq. \eqref{ptrh0H01} for Yukawa Lagrangian in Eq. \eqref{quark}, we derive easily that 
\begin{align}
	\label{eq:h0ff}
	\mathcal{L}^{hff} & = \sum_{f=\ell,u,d}\frac{\sqrt{2}}{k} \overline{f'_{L}} \hat{M}_f f'_{R}h +\mathrm{H.c.}\simeq  \sum_{f=\ell,u,d}\frac{g}{\sqrt{2}m_W} \overline{f_{L}} M_f f_{R} h +\mathrm{H.c.},
\end{align}
where $k=246=g/(\sqrt{2}m_W)$, and the transformation in Eq. \eqref{eq:h0ff} is based on discussion relating to Eq. \eqref{eq:Mud}. Therefore, the SM-like Higgs couplings with charged fermions can be identified with the SM results.

Similarly, the original and mass  states of the singly charged Higgs bosons have the following relations
\begin{align} 
&\begin{pmatrix}
\phi^\pm_1\\
\phi^\pm_2\\
\end{pmatrix}
=\begin{pmatrix}
	-s_{\beta } & c_{\beta } \\
	c_{\beta } & s_{\beta } \\
\end{pmatrix}	\begin{pmatrix}
G^{\pm}_W\\
H^{\pm}_1\\
\end{pmatrix}, \crn
&  H^\pm_2\simeq \delta_L^{\pm}, \;  G_{W_2}^\pm\simeq \delta_R^{\pm},
\end{align}
where $G_{W_2}^\pm$ is massless, corresponding to the Goldstone boson eaten up by $W_2^\pm$,   and the remaining squared masses of singly charged Higgs bosons are 
\begin{align}
\label{eq_m2hpm}
m^2_{H^\pm_1}=	 \frac{\alpha_3 v^2_R}{2\left( 2s^2_{\beta} -1\right)},\; m^2_{H^\pm_2}=\frac{1}{2} \left(\rho_3-2 \rho_1\right) v_R^2. 
\end{align}
Besides that,  two components ($\delta^{++}_L,\delta^{ ++}_R$) are also physical states  with the following masses   
\begin{align}
\label{eq_mh2pp}
m^2_{H^{\pm \pm}_1}=\frac{1}{2} v^2_R(\rho_3 -2\rho_1),\; \quad  m^2_{H^{\pm \pm}_2}=2 v^2_R\rho_2. 
\end{align}

\section{\label{sec:coupling} Couplings and analytic formulas  involved with loop-induced Higgs decays}

\subsection{Couplings}
From the above  Higgs potential and the discussion on the masses and mixing of Higgs bosons, all  Higgs self-couplings of  $h$ giving one-loop contributions to the decays $h \rightarrow \gamma\gamma, Z\gamma$ can be derived analytically. From the general notations in  the interacting Lagrangian: $-V_h\to \mathcal{L}_{hSS}=\sum_{S_i,S_j=H^\pm_{1,2},H^{\pm\pm}_{1,2}} (-\lambda_{hS_{ij}} hS_i^{Q_S}S_j^{-Q_S} +\mathrm{h.c.}) +\dots$, the  Feynman rule $-i\lambda_{hSS}$ corresponds to the vertex $hSS$.  All non-zero factors $\lambda_{hSS}$ are given in  Table \ref{table_h0coupling}.
\begin{table}[h]
	\centering
	\begin{tabular}{|c|c|}
\hline 
	Vertex		& Coupling: $\lambda_{hSS}$ \\ 
\hline 
		$\lambda_{h H^+_1 H^-_1}$	&$2 \left\lbrace c_\beta ^4 \lambda_1 +2 c_\beta^2s_\beta^2 [\lambda_1 - 2 (2 \lambda_2 + \lambda_3)] + \lambda_1 s_\beta^4\right\rbrace  v$ \\
\hline 
		$\lambda_{h H^+_2 H^-_2}$	& $\frac{1}{2}(2\alpha_1+\alpha_3+8\alpha_2 c_\beta s_\beta)v$ \\ 
\hline 
		$\lambda_{h H^{++}_1 H^{--}_1}$	&$\left[ \alpha_1+s_\beta(4\alpha_2 c_\beta +\alpha_3 s_\beta)\right] v$ \\ 
\hline 
		$\lambda_{h H^{++}_2 H^{--}_2}$	&  $\left[  \alpha_1 +s_\beta(4\alpha_2 c_\beta +\alpha_3 s_\beta)\right] v$ \\
\hline 
		$\lambda_{h H^{++}_1 H^{--}_2}$	&  $   (\alpha_6+\alpha_4 c_\beta s_\beta)(-1+2s_\beta^2) \frac{v}{s_{\beta}} $ \\
\hline 
		$\lambda_{h H^{++}_2 H^{--}_1}$	&$   (\alpha_6+ \alpha_4 c_\beta s_\beta)(-1+2s_\beta^2) \frac{v}{s_{\beta}} $ \\
\hline
\end{tabular} 
	\caption{Feynman rules for  the SM-like  Higgs boson couplings with charged Higgs bosons} \label{table_h0coupling}
\end{table}
We note that the vertex factors in Table \ref{table_h0coupling} are derived following the general notation defined in Ref. \cite{Hue:2017cph}, so that we can use the analytic formulas to compute the partial decay widths $h\to \gamma \gamma, Z \gamma$ in the MLRSM mentioned in this work.  

The couplings of $h$ with SM fermions can be determined using the Yukawa Lagrangians given in Eqs. \eqref{quark}, where  the Feynman rule is $-i\left(Y_{h\bar{f}fL}P_L + Y_{h\bar{f}fR}P_R\right)$ for each vertex $h\bar{f}f$. Because this model does not have exotic charged fermions and the couplings of SM leptons to neutral Higgs/gauge bosons ($g_{h\bar{f}f}, g_{Z\bar{f}f}$) are defined as in the SM \cite{Dedes:2019bew, Dawson:2018pyl, Cao:2018cms}, we will use the SM results for   one-loop fermion contributions to the decay amplitudes $h\to Z\gamma,\gamma \gamma$.  

The Higgs-gauge boson couplings giving one-loop contribution to the decays $h\rightarrow Z\gamma,\gamma\gamma$ are derived from  the  kinetic Lagrangian of the Higgs bosons, namely
\begin{align}\label{eq_lkHiggs}
\mathcal{L}^{H}_{\mathrm{kin}}&= \mathcal{L}_\Phi +	\mathcal{L}_{\Delta_L} +\mathcal{L}_{\Delta_R} \crn
&=  \sum_{i,j=1}^2 g_{\mu\nu} g_{hW_iW_j} hW_i^{-\mu} W_j^{+\nu} \crn 
+&\sum_{S_i,j}\left[ -ig^*_{hSW_j}W_j^{-\mu}\left( S_i^{+Q}\partial_{\mu}h -h\partial_{\mu}S_i^{+Q} \right) + ig_{hSW_j}W_j^{+\mu}\left( S_i^{-Q}\partial_{\mu}h -h\partial_{\mu}S_i^{-Q} \right) \right]\crn 
+&\sum_{S_i,S_j}ig_{ZS_iS_j}Z^{\mu} \left( S_i^{-Q}\partial_{\mu}S_j^{Q} -S_j^{Q}\partial_{\mu}S_i^{-Q} \right) \crn 
+&\sum_{S_i,j}\left[ ig_{ZW_jS_i}Z^{\mu} W_j^{+\nu}S_i^{-Q} g_{\mu\nu} + ig^*_{ZW_jS_i}Z^{\mu} W_j^{-\nu}S_i^{Q} g_{\mu\nu}\right] \crn
+ &\sum_{S_i} ie Q A^{\mu}\left( S_i^{-Q}\partial_{\mu}S_i^{Q} -S_i^{Q}\partial_{\mu}S_i^{-Q} \right) +..., 
\end{align}
where $S_i,S_j=H^{\pm}_{1,2}, H^{\pm\pm}_{1,2}$ denote charged Higgs  bosons in the MLRSM.  The  Feynman  rules for the $h$ couplings to at least one charged gauge boson are shown in Table \ref{table_HGcoupling}.  
\begin{table}[h]
	\centering
	\begin{tabular}{|c|c|}
\hline
Vertex	& Coupling \\ 
\hline
$g_{h W_1^+W_1^-}$&$\frac{1}{2} \left\lbrace -4g_L g_R s_\beta c_\beta  s_{\xi} c_{\xi}   +  (c_{\xi}^2 g_L^2 + g_R^2 s_{\xi}^2) \right\rbrace v$\\ 
\hline
		$g_{h W_2^+W_2^-}$&$\frac{1}{2}  \left\lbrace 4g_L g_R s_\beta c_\beta  s_{\xi} c_{\xi}   +  (s_{\xi}^2 g_L^2 + g_R^2 c_{\xi}^2) \right\rbrace v$\\
\hline
		$g_{h W_1^+W_2^-}$&$\frac{1}{2}\left\lbrace s_{\xi} c_{\xi} (-g_L^2 + g_R^2)  + 2  g_L g_R s_\beta c_\beta (-c_{\xi}^2 + s_{\xi}^2)\right\rbrace v$\\
\hline
		$g_{h W_2^+W_1^-}$&$ \frac{1}{2} \left\lbrace s_{\xi} c_{\xi} (-g_L^2 + g_R^2)  + 2  g_L g_R s_\beta c_\beta (-c_{\xi}^2 + s_{\xi}^2)\right\rbrace v$\\
\hline
		$g_{h H_1^{-}W_1^{+}}$&$ \frac{1}{2} g_R s_{\xi}(c_\beta^2 - s_\beta^2) $ \\
\hline
		$g_{h H_1^{-}W_2^{+}}$ &  $\frac{1}{2} g_R c_{\xi} (c_\beta^2 - s_\beta^2)$\\
\hline
\end{tabular}
\caption{Feynman rules for couplings of the SM-like Higgs boson to charged Higgs and  gauge bosons. \label{table_HGcoupling} }
\end{table}
The momenta appearing in the vertex factors are $\partial_{\mu}h\rightarrow -ip_{0\mu}h$ and $\partial_{\mu}S_{i,j}\rightarrow -ip_{\mu}S_{i,j}$, where  $p_0$, $p_{\pm}$  are incoming momenta.

The Feynman rules for $Z$ couplings  to  charged Higgs and gauge bosons in Eq. \eqref{eq_lkHiggs}  are given in Table \ref{table_Z1A}. The couplings  $g_{ZH_i^{+,++}H_j^{-,--}},g_{ZW_2^+H_{i,j}^-} $ are zero in the MLRSM. 
\begin{table}[h]
	\centering
	\begin{tabular}{|c|c|}
\hline
Vertex	& Coupling\\
\hline
$g_{ZH_1^+H_1^-}$	&$\frac{1}{2}(p_{+}-p_{-})[c_{z_2} c_{z_3} g_L - g_R (c_R c_{z_3} s_{z_2} + s_R s_{z_3})] $ \\	
\hline
		$g_{ZH_2^+H_2^-}$	&$g' (p_{+}-p_{-}) (-c_{z_3} s_R s_{z_2} + c_R s_{z_3})$ \\
\hline
		$g_{ZH_1^{++}H_1^{--}}$	&$ (p_{++}-p_{--} ) (c_{z_2} c_{z_3} g_L - c_{z_3} g' s_R s_{z_2} + c_R g' s_{z_3})$ \\
\hline
		$g_{ZH_2^{++}H_2^{--}}$	&$( p_{++}-p_{--}) (c_R c_{z_3} g_R s_{z_2} + c_{z_3} g' s_R s_{z_2} - c_R g' s_{z_3} + g_R s_R s_{z_3})$ \\
\hline
		$g_{ZW_1^{+}H_1^-}$& $\frac{1}{2} c_{z_2} c_{z_3} g_L g_R (c_\beta^2 - s_\beta^2) s_{\xi} v$\\
\hline
		$g_{ZW_1^{+}H_2^-}$&$\frac{1}{2} c_{\xi} c_{z_2} c_{z_3} g_L g_R (c_\beta^2 - s_\beta^2) v$\\
\hline
\end{tabular}
\caption{Feynman rules of couplings of  $Z$ to charged Higgs and gauge bosons.} Notations $p_+$ and $p_-$ are incoming momenta. \label{table_Z1A} 
\end{table}

The triple gauge couplings of  $Z$ and photon to $W^\pm_{1,2}$ are derived from the  kinetic Lagrangian of the non-Abelian gauge bosons
\begin{align} \label{eq_Lg}
\mathcal{L}^k_g= -\frac{1}{4}\sum_{a=1}^3 \left[ F^a_{L\mu\nu}F^{a\mu\nu}_{L} + F^a_{R\mu\nu}F^{a\mu\nu}_{R}\right],
\end{align}
where 
$F^a_{L,R\mu \nu}=\partial_{\mu}W^a_{L,R\nu} -\partial_{\nu}W^a_{L,R\mu} +g_{L,R} \epsilon^{abc}W^b_{L,R\mu}W^c_{L,R\mu}$, and 
$\epsilon^{abc}$ $(a,b,c=1,2,3)$ are the $SU(2)$ structure constants.  The respective $Z$ couplings  to $W^\pm_{1,2}$ are  included in the following part: 
\begin{align} \label{LZWW}
\mathcal{L}^{ZW^{+}W^{-}}\subset - g_L\epsilon^{abc}(\partial_\mu W^a_{L\nu})W^{b\mu}_LW^{c\nu}_L - g_R\epsilon^{abc}(\partial_\mu W^a_{R\nu})W^{b\mu}_RW^{c\nu}_R.
\end{align}
Then, the vertex factors corresponding to particular couplings  are defined as 
\begin{align}\label{eq_ZAgg}
\mathcal{L}^g_D \rightarrow& -g_{ZW_iW_j}Z^{\mu}(p_0)W^{+\nu}_i(p_+) W^{-\lambda}_j(p_-)\times \Gamma_{\mu\nu\lambda}(p_0,p_+,p_-), \crn 
&-eA^{\mu}(p_0)W_i^{+\nu}(p_+)W_i^{-\lambda}(p_-)\times \Gamma_{\mu\nu\lambda}(p_0,p_+,p_-),
\end{align}
where $\Gamma_{\mu\nu\lambda}(p_0,p_+,p_-) \equiv g_{\mu\nu}(p_0-p_+)_{\lambda} +g_{\nu\lambda}(p_+ -p_-)_{\mu} +g_{\lambda\mu}(p_--p_0)_{\lambda}$, and $i,j=1,2$. The photon always couples to two identical particles as the consequence of the Ward Identity  \cite{Hue:2023rks},  see the second line of Eq. \eqref{eq_ZAgg}. 
The non-zero factors for  triple couplings of  $Z$  with charged gauge bosons are collected in Table~\ref{table_3gaugcoupling}.    
\begin{table}[h]
	\centering
	\begin{tabular}{|c|c|}
\hline
Vertex	& Coupling \\ 
\hline
$g_{ZW^{+\nu}_1W^{-\lambda}_1}$& $\frac{-e}{s_{z_2}c_Rc_{z_2} } \left[c_Rc_{z_3} (-c^2_{z_2} + s^2_\xi ) +s_Rs_{z_3}s_{z_2}s^2_{\xi} \right]$\\
\hline
		$g_{ZW^{+\nu}_2W^{-\lambda}_2}$& $ \frac{e}{s_{z_2}c_Rc_{z_2} }\left[ c_Rc_{z_3} (s^2_{\xi}  -s^2_{z_2}) -s_Rs_{z_3}s_{z_2} c^2_{\xi} \right] $ \\
\hline
		$g_{ZW^{+\nu}_1W^{-\lambda}_2}$& $-\frac{e s_\xi c_\xi}{s_{z_2}c_Rc_{z_2} }\left[ c_{z_3}c_R + s_Rs_{z_3}s_{z_2}\right]$\\
\hline
		$g_{ZW^{+\nu}_2W^{-\lambda}_1}$& $-\frac{e s_\xi c_\xi}{s_{z_2}c_Rc_{z_2} }\left[ c_{z_3}c_R + s_Rs_{z_3}s_{z_2}\right]$\\
\hline
\end{tabular}
\caption{Feynman rules for triple  gauge couplings  relating with the decay $h\rightarrow Z\gamma$. \label{table_3gaugcoupling} }
\end{table}

To end this section, we emphasize that all couplings determined in this section do not use the assumption $k_1\gg k_2$, equivalently $t_{\beta}\gg1$ used in Ref. \cite{Zhang:2007da,Lee:2017mfg}.

\subsection{Partial decay widths and signal strengths of the  decays $h\rightarrow Z\gamma,\gamma\gamma$}
\begin{figure}[ht]
	\centerline{\includegraphics[width=4.0in]{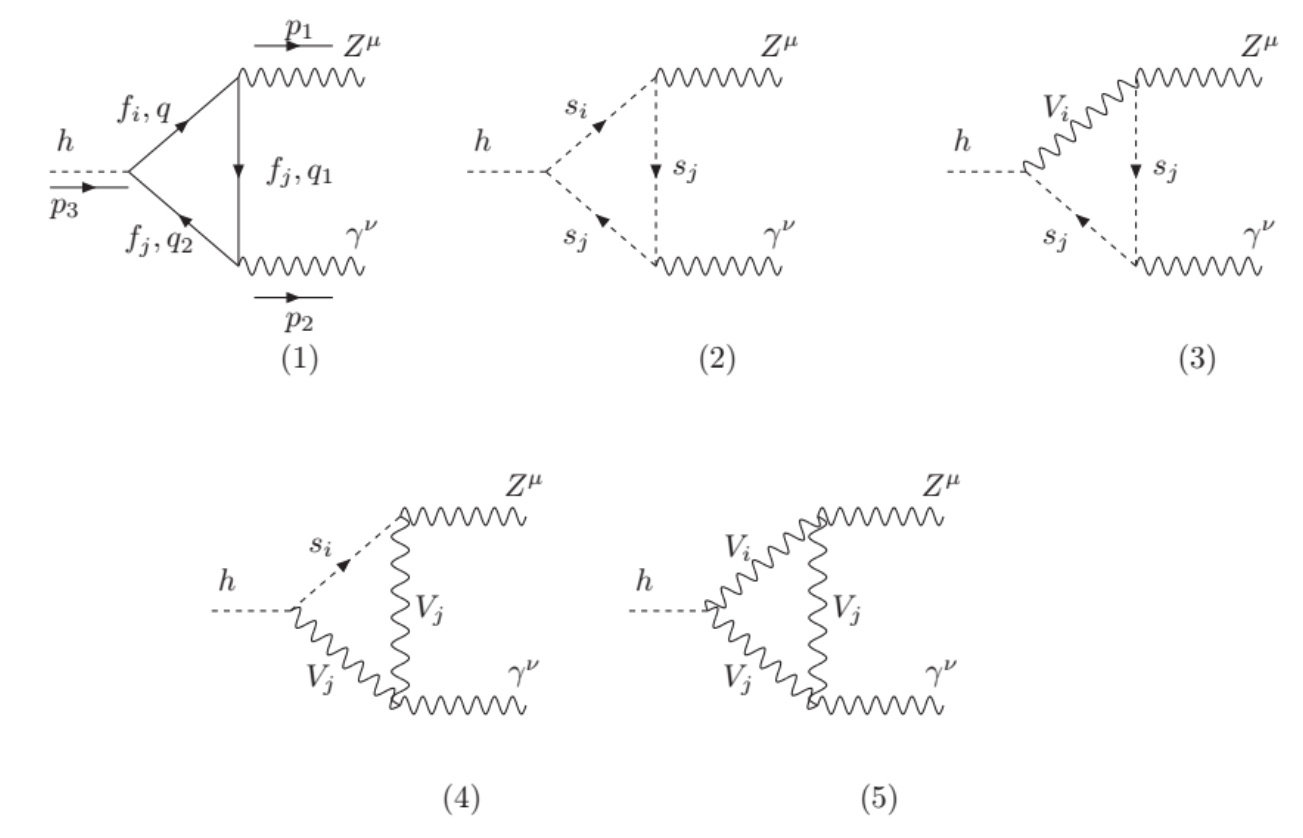}}
\vspace*{8pt}
	\caption{ One-loop three-point Feynman diagrams contributing to the decay $h\rightarrow\,Z\gamma$ in the unitary gauge, where $f_{i,j}$ are the SM fermions,  $s_{i,j}=H^{\pm}_{1,2},H^{\pm \pm}_{1,2}$, $V_{i,j}=W_1^{\pm}, W_2^{\pm }$.\protect \label{fig_hzgaDiagram}}
\end{figure}

In the  MLRSM framework, one-loop three-point Feynman diagrams giving contributions to the decay amplitude $h\rightarrow\, Z\gamma$ are shown  in Fig.~\ref{fig_hzgaDiagram},
where  the unitary gauge is applied to determine the gauge boson contributions. The fermion contributions    to amplitude of the decay $h \to Z \gamma$   coincide with the SM results calculated in {Ref. \cite{Dedes:2019bew, Dawson:2018pyl}}. Using the general calculation introduced in Ref. \cite{Hue:2017cph}, we can write these contributions as follows
\begin{equation}\label{eq_F21fLR}
F_{21,f}^{\mathrm{LR}}=F_{21,f}^{\mathrm{SM}}=
\sum_{f_i=e_i,u_i,d_i}F_{21,f_i}^{\mathrm{SM}},
\end{equation}
where all form factors $F_{21,f_i}^{\mathrm{SM}}$ are written in terms of the Passrino-Veltman (PV) notations \cite{Passarino:1978jh}. 
 
Similarly, the contribution from the charged Higgs bosons can be given as 
\begin{align}
F^{\mathrm{LR}}_{21,S}&= F_{21,H^+_1} + F_{21,H^+_2} + F_{21,H^+_{122}} + F_{21,H^+_{211}} \crn 
&+ F_{21,H^{++}_1} +F_{21,H^{++}_2} +F_{21,H^{++}_{122}} +F_{21,H^{++}_{211}}.\label{eq_F21sLR}
\end{align}
The  charged gauge boson contributions $W^\pm_{1,2}$ to the $h \to Z \gamma$ amplitude   are
\begin{align}
F^{\mathrm{LR}}_{21,V}&= F^{\mathrm{LR}}_{21,W^+_1} + F^{\mathrm{LR}}_{21,W^+_2} + F^{\mathrm{LR}}_{21,W^+_{122}} + F^{\mathrm{LR}}_{21,W^+_{211}} .\label{eq_F21SVLR}
\end{align}
Similarly, the contribution from charged Higgs and  gauge boson arising  from two diagrams 3 and 4 in Fig. \ref{fig_hzgaDiagram} can be given as 
\begin{align}
F^{\mathrm{LR}}_{21,VS}&= F_{21,WSS}^{\mathrm{LR}}+F_{21,SWW}^{\mathrm{LR}}, \label{eq_F21SVLR1}
\end{align}
where 
\begin{align}
F^{\mathrm{LR}}_{21,WSS}&= F_{21,W^+_1H^+_1H^+_1}^{\mathrm{LR}} + F_{21,W^+_1H^+_2H^+_2}^{\mathrm{LR}} + F_{21,W^+_2H^+_1H^+_1}^{\mathrm{LR}} + F_{21,W^+_2H^+_2H^+_2}^{\mathrm{LR}}  , \label{eq_F21VSSLR} \\
F^{\mathrm{LR}}_{21,SWW}&= F_{21,H^+_1W^+_1W^+_1}^{\mathrm{LR}} + F_{21,H^+_1W^+_2W^+_2}^{\mathrm{LR}} + F_{21,H^+_2W^+_1W^+_1}^{\mathrm{LR}} + F_{21,H^+_2W^+_2W^+_2}^{\mathrm{LR}}. \label{eq_F21SVVLR}
\end{align}

Now, the $h\to Z \gamma$ partial decay width is~\cite{Gunion:1989we, Degrande:2017naf}
\be
\Gamma^{\mathrm{LR}}(h\rightarrow  Z\gamma)=\frac{m_h^3}{32\pi}
\times \left(1-\frac{m_Z^2}{m_h^2}\right)^3 |F^{\mathrm{LR}}_{21}|^2,
\label{GaHZga1}
\ee
where the scalar factors  $F^{\mathrm{LR}}_{21}$ is derived as follows  \cite{Hue:2017cph}
\begin{align}
F^{\mathrm{LR}}_{21}&= F^{\mathrm{LR}}_{21,f} + F^{\mathrm{LR}}_{21,S} +F^{\mathrm{LR}}_{21,V}+  F^{\mathrm{LR}}_{21,VS}.
\label{eq_F21331be}
\end{align} 
We note that $F^{\mathrm{LR}}_{21,VS}$ were omitted  in some previous works~\cite{Yue:2013qba,Martinez:1989kr, Martinez:1990ye} because it was expected to be much smaller than the contributions from the SM and are still far from the sensitivity of the recent experiments. However, since collider sensitivities have recently been improved and new scales have been established, these contributions are necessary.
The branching ratio   $\mathrm{Br}^{\mathrm{LR}}(h\rightarrow Z\gamma)$ in the MLRSM framework is
\begin{equation} \label{eq_brZgaLR}
\mathrm{Br}^{\mathrm{LR}}(h\rightarrow Z\gamma)= \frac{	\Gamma^{\mathrm{LR}}(h\rightarrow Z\gamma)}{	\Gamma^{\mathrm{LR}}_h},
\end{equation}
where $\Gamma^{\mathrm{LR}}_{h}$ is the total decay width of the SM-like Higgs boson
$h$ ~\cite{Gunion:1989we, Degrande:2017naf}. 
Although  there are available  experimental measurements of the SM-like Higgs boson productions and decays   \cite{Khachatryan:2016vau},  we focus  only on  the Higgs production through the gluon fusion process $ggF$ at LHC, in which  the respective signal strength predicted by two models  SM and MLSM are equal. Then the signal strength corresponding to the decay mode $h\to Z \gamma$ predicted by the MLRSM is:
\begin{equation}\label{eq_muZga}
\mu_{Z\gamma}^{\mathrm{LR}}\equiv \frac{\mathrm{Br}^{\mathrm{LR}}(h\rightarrow Z\gamma)}{\mathrm{Br}^{\mathrm{SM}}(h\rightarrow Z\gamma)},
\end{equation}
where  $ \mathrm{Br}^{\mathrm{SM}}(h\rightarrow Z\gamma)$ is the SM branching ratio of the decay $h \to Z \gamma$. The recent $ggF\to h\to Z\gamma$ singal strength  is  $\mu_{Z\gamma} =2.4 \pm 0.9$ at $2.7\sigma$ (standard deviation) \cite{CMS:2022ahq, CMS:2023mku}. 

Similarly, the  partial decay width and signal strength of the decay $h\rightarrow \gamma\gamma$ can be calculated as \cite{Degrande:2017naf, Hue:2017cph}
\begin{align}
\label{eq_Gah2gamma}
\Gamma^{\mathrm{LR}}(h\rightarrow  \gamma\gamma) =&\frac{m_h^3}{64\pi}
\times |F^{\mathrm{LR}}_{\gamma \gamma}|^2 ,
\crn \mu_{\gamma\gamma}^{\mathrm{LR}}\equiv& \frac{\Gamma^{\mathrm{LR}}(h\rightarrow \gamma\gamma)}{\Gamma^{\mathrm{SM}}(h\rightarrow \gamma\gamma)}, 
\end{align}
where 
\begin{align}\label{eq_F2gamma}
	F^{\mathrm{LR}}_{\gamma\gamma}&=\sum_{f}F^{\mathrm{LR}}_{\gamma \gamma,f} + \sum_{s}F^{\mathrm{LR}}_{\gamma \gamma,s} +\sum_{v}F^{\mathrm{LR}}_{\gamma \gamma,v},
\end{align}
and 
\begin{align}
\label{eq_FLRgaga}
F_{\gamma\gamma,f}^{\mathrm{LR}}= & F_{\gamma\gamma,f}^{\mathrm{SM}}=
\sum_{f_i=e_i,u_i,d_i}F_{\gamma\gamma,f_i}^{\mathrm{SM}},
\crn 	F^{\mathrm{LR}}_{\gamma\gamma,S} =& F_{\gamma\gamma,H^+_1} + F_{\gamma\gamma,H^+_2} + F_{\gamma\gamma,H^{++}_1} +F_{\gamma\gamma,H^{++}_2},
\crn F^{\mathrm{LR}}_{\gamma\gamma,V}= & F_{\gamma\gamma,W^+_1} + F_{\gamma\gamma,W^+_2}. 
\end{align}
Here we have used the notations that \cite{Hung:2019jue} 
\begin{align}
	\label{eq_Fgaga}
	F^{\mathrm{SM}}_{\gamma \gamma,f_i}&=- \frac{e^2\,Q^2_{f_i}\,N_c}{2\pi^2}\left( m_f Y_{h\bar{f}fL}\right)  \left[4  X_2 +C_0\right], \crn 
	F_{\gamma \gamma,s}&= \frac{e^2\,Q^2_s \lambda_{hss}}{2\pi^2} X_2, \crn
	F_{\gamma \gamma,v}&=\frac{e^2\,Q^2_V\,g_{hvv}}{4\pi^2} \times\left\{ \left( 6+ \frac{m_h^2}{m_V^2}\right) X_2   +  4C_0 \right\},
\end{align}
where $X_2=C_{12} +C_{22} +C_2$  and $C_{0,i,ij}\equiv C_{0,i,ij}(0,0,m_h^2; m_x^2, m_x^2, m_x^2)$ are PV functions \cite{Passarino:1978jh} with $x=f,s,v$ implying  fermions, charged Higgs and gauge bosons, respectively. Particular forms given in  Eq. \eqref{eq_Fgaga} are defined precisely in Ref. \cite{Hung:2019jue}. In the following section, the numerical results will be evaluated using LoopTools \cite{Hahn:1998yk}. 

\section{\label{sec:numerical} Numerical discussions}
\subsection{\label{setuppa} Setup parameters}

In this section,  there are following  quantities  fixed from experiments~\cite{Tanabashi:2018oca}:    $m_h=125.38$ GeV,  $m_W$, $m_Z$,  well-known fermion masses,  $v\simeq 246$ GeV, the $SU(2)_L$ gauge coupling $g_2\simeq 0.651$,  $\alpha_{\mathrm{em}}=1/137$, $e=\sqrt{4\pi\alpha_{\mathrm{em}}}$, $s^2_W=0.231$. 

The unknown Higgs self-couplings of the MLRSM are  $\rho_{1,2,3,4}$, $\alpha_{1,2,3,4,5,6}$, $\lambda_{2,3,4}$. The dependent parameter  $\lambda_1$ is given  by Eq.~\eqref{eq_fla1}. 
Some Higgs self-couplings are expressed as functions of  the  heavy Higgs boson masses, namely
\begin{align}
\label{eq_m2sx}	
m^2_{H^0_1}& = m^2_{A_1}= m^2_{H^\pm_1} =\frac{\alpha_3  v_R^2}{2 \left( 2s_{\beta}^2 -1\right)}, 
\crn  m^2_{H^0_2} &= m^2_{H^{\pm }_2}= m^2_{H^{\pm\pm }_1}= m^2_{A_2}= \frac{ v_R^2 }{2} \left( -2 \rho_1 +\rho_3\right),\; 
\crn  m^2_{H^0_3}&=2\rho_1 v_R^2, \;  m^2_{H^{\pm\pm}_2} =2\rho_2 v_R^2.
\end{align}
Choosing the mass of  $m_{H^+_1}$, $m_{H^+_2}$, and $m_{H^{++}_1}$ as free parameters we get
\begin{align}
\label{eq_lax}
\alpha_1&= \frac{2
	m_{H_2^+}^2}{\left(t_{\beta }^2+1\right) v_R^2}, \;
\alpha_2= -\frac{t_{\beta } m_{H_2^+}^2}{\left(t_{\beta }^2+1\right) v_R^2}, \; 
\alpha_3= \frac{2 \left(t_{\beta }^2-1\right) m_{H_2^+}^2}{\left(t_{\beta }^2+1\right) v_R^2}  ,\;
\crn \alpha_4&= -\frac{2 \alpha_6}{t_{\beta }}, \; \rho_2=\frac{m_{H_2^{\text{++}}}^2}{2 v_R^2}, \; \rho_3= 2 \rho_1+\frac{2 m_{H_1^+}^2}{v_R^2}. 
\end{align}
The other free parameters are
$\lambda_{2,3,4}$, $\rho_1= m^2_{H^0_3}/(2 v_R^2)>0$,  the mixing angle and the gauge bosons masses will be at the orders of $\mathcal{O}(v^2/v_R^2)$
\begin{align}
\label{eq_input}
g_R&=g_L=g_2=e/s_W,\; g'=\frac{e}{\sqrt{1-2s^2_W}},
\crn
s_{\xi}&=- \frac{s_{\beta} c_{\beta} v^2}{v_R^2}\ll1 ,\; c_{\xi}=1+\mathcal{O}\left( \frac{v^4}{v_R^4}\right) \simeq 1,
\crn 
s_R&= \frac{\sqrt{1-2s_W^2}}{c_W},\; c_R = t_W, 
\crn 
s_{z_2}&=s_W ,\; c_{z_2} = c_W, 
\crn  
s_{z_3}&= \frac{t_W^2\sqrt{1-2s_W^2}v^2}{4c_W^2 v_R^2}\ll1 ,\; c_{z_3} =1+\mathcal{O}\left( \frac{v^4}{v_R^4}\right)\simeq 1.
\end{align}
 We note here that the relations given in Eq. \eqref{eq_input} are consistent with  the SM because   the two couplings  $hW^+W^-$ and $ZW^+W^-$  are consistent with the SM predictions. 
 
Apart from the limit $g_L=g_R$ chosen in Eq. \eqref{eq_input}, various discussions for  the more general case $g_R\neq g_L$, which  showed that this ratio is allowed in the follows range \cite{Lindner:2016lpp, Chauhan:2018uuy}:
 	\begin{equation}
 		\label{eq:gRL}
 		0.65\leq g_R \leq 1.6,
 	\end{equation}
 	where the lower bound $v_R>10$ TeV. 	

A recent study showed a lower bound of $m_{W_R}>5.5$ TeV is still allowed \cite{Dekens:2021bro}, which gives $v_R\ge 17$ TeV in this case. On the other hand, the constraints of $t_{\beta}\geq 1.2$ is allowed, while no lower bounds of charged Higgs masses were given, especially in the limit of the phase $\alpha$ given in Eq. \eqref{vevhigg1} is zero.  Various of works discuss the constraints of Higgs masses indirectly \cite{Maiezza:2016ybz}, or directly from LHC for doubly charged Higgs bosons \cite{ATLAS:2022pbd}. The lower bounds are $m_{H^{\pm\pm}}\geq 1080 $ GeV. Theoretical constraints was discussed in Ref. \cite{Chakrabortty:2016wkl} for Higgs self couplings satisfying unitarity bounds and vacuum stability criteria, which will be applied in our numerical investigation. 

Based on the above discussion for  investigating the significant strengths of the two decays $h \to\gamma\gamma,  Z \gamma$, the  values of unknown independent parameters we choose here will be scanned in the following ranges:
 \begin{align}
 \label{eq:scanningrange}
& m_{H^+_1},\; m_{H^+_2}, m_{H^{++}_2}\in [1,\; 20]\;\mathrm{ TeV},\; v_R \in [20,\; 60] \;\mathrm{TeV},\; t_{\beta} \in [1.2,\; 30],
\crn& \lambda_{2,3,4} \in [-10,10],
\end{align}
where the Higgs self-couplings satisfy all theoretical constraints discussed on Ref.   \cite{Chakrabortty:2016wkl}.

\subsection{Results and discussions} 
\begin{figure}[ht]
\centerline{\includegraphics[width=3.0in]{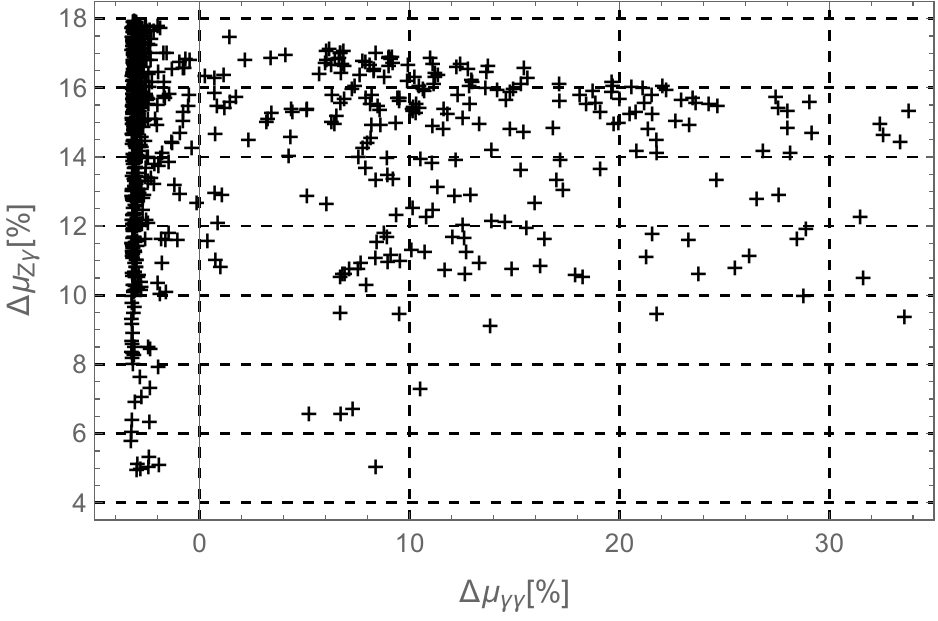}}
\vspace*{8pt}
	\caption{Correlations between $\Delta\mu_{Z\gamma}^{\mathrm{LR}}$ and $\Delta\mu_{\gamma \gamma}^{\mathrm{LR}}$  with $g_L=g_R$.\protect\label{fig_muZga-2ga}}
\end{figure}

To express the differences of the prediction between the SM and the MLRSM, we define a quantity $\Delta\mu_{Z \gamma}$  as in Ref.~\cite{Hung:2019jue}   
\begin{align}\label{eq_demu}
\Delta{\mu}^{\mathrm{LR}}_{Z \gamma}\equiv \left(\mu^{\mathrm{LR}}_{Z \gamma}-1\right)\times 100\%,
\end{align}  
which is constrained by recent experiments  $\Delta \mu_{Z\gamma} =1.4 \pm 0.9$  \cite{CMS:2022ahq, CMS:2023mku}, implying the following $1\sigma$ deviation:
\begin{equation}
	\label{eq_demuZgaexp}
50\% \le \Delta{\mu}^{\mathrm{LR}}_{Z \gamma} \le 230 \%. 
\end{equation}
The $1\sigma$ constraint from $h\to \gamma \gamma$ decay originating from $ggF$ fusion is defined as $\Delta \mu^{\mathrm{LR}}_{\gamma\gamma} \equiv ( \mu^{\mathrm{LR}}_{\gamma\gamma}-1) \times 100\% $, leading to the  respective $2\sigma$ deviation as follows
\begin{align}
	\label{eq_dmu2ga}
-12\%<\Delta \mu^{\mathrm{LR}}_{\gamma\gamma}<38\%.
\end{align}
The numerical results we discuss in the following will always  satisfy this constraint.  We have checked numerically that the MLRSM always contains regions of the parameter space  that both values of $\Delta{\mu}_{Z\gamma},\Delta{\mu}_{\gamma\gamma} \to 0$, implying the  consistency  with the SM results. Considering the special case of $g_L=g_R$, we discuss  firstly on  the dependence  of $\Delta \mu^{\mathrm{LR}}_{Z\gamma}$ on $\Delta \mu^{\mathrm{LR}}_{\gamma \gamma}$,  which is  illustrated in Fig.~\ref{fig_muZga-2ga}.
We just focus in the region satisfying $|\Delta\mu^{\mathrm{LR}}_{Z\gamma}|\ge 5\%$  in order to collect interesting points that may support the $1\sigma$ range given in Eq. \eqref{eq_demuZgaexp}
It can be seen that $\Delta\mu_{Z\gamma}^{\mathrm{LR}}$ is constrained strictly by   $\Delta\mu_{\gamma\gamma}^{\mathrm{LR}}$, i.e., $\Delta\mu_{Z\gamma}^{\mathrm{LR}}\leq 18\%$ in the range of  $2\sigma$ deviation given in Eq. \eqref{eq_dmu2ga}. It is noted that negative values of  $\Delta\mu_{\gamma\gamma}^{\mathrm{LR}}<0$ can give large $\Delta\mu_{Z\gamma}^{\mathrm{LR}}$ than the positive ones. Largest values of  $\Delta\mu_{Z\gamma}^{\mathrm{LR}}$  is still much smaller than the $1\sigma$ deviation given by recent experimental data.

We comment here a point that the future sensitivities are  $|\Delta\mu_{\gamma\gamma}|\leq4\%$, and $|\Delta\mu_{Z\gamma}|\leq23\%$, respectively \cite{Cepeda:2019klc}. In the model under consideration,    large  values of $\Delta\mu_{Z\gamma}>23\%$ are not allowed with $g_L=g_R$.

\begin{figure}[ht]
	\centerline{
		\begin{tabular}{cc}
			\includegraphics[width=6cm]{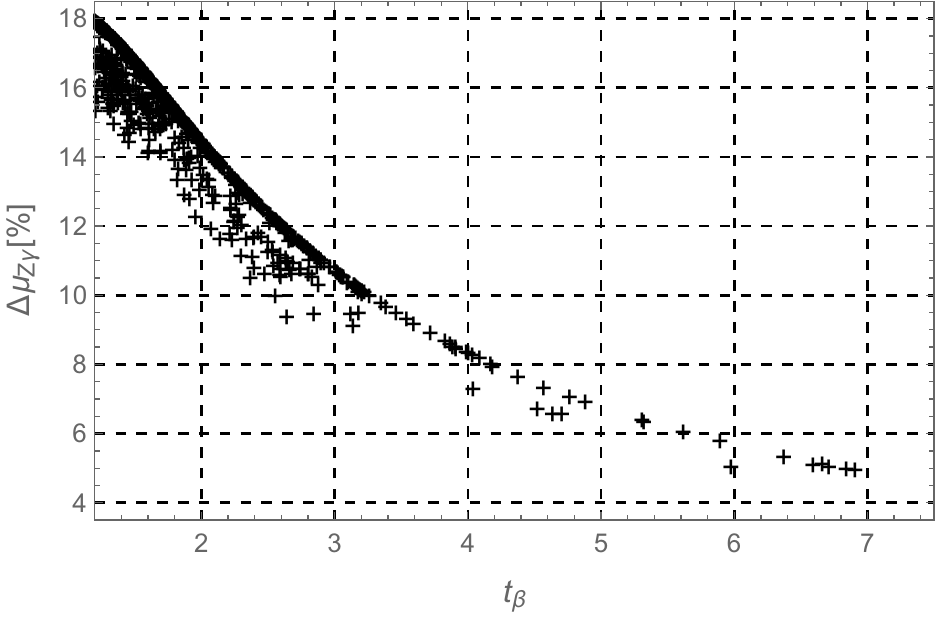}
			\includegraphics[width=6cm]{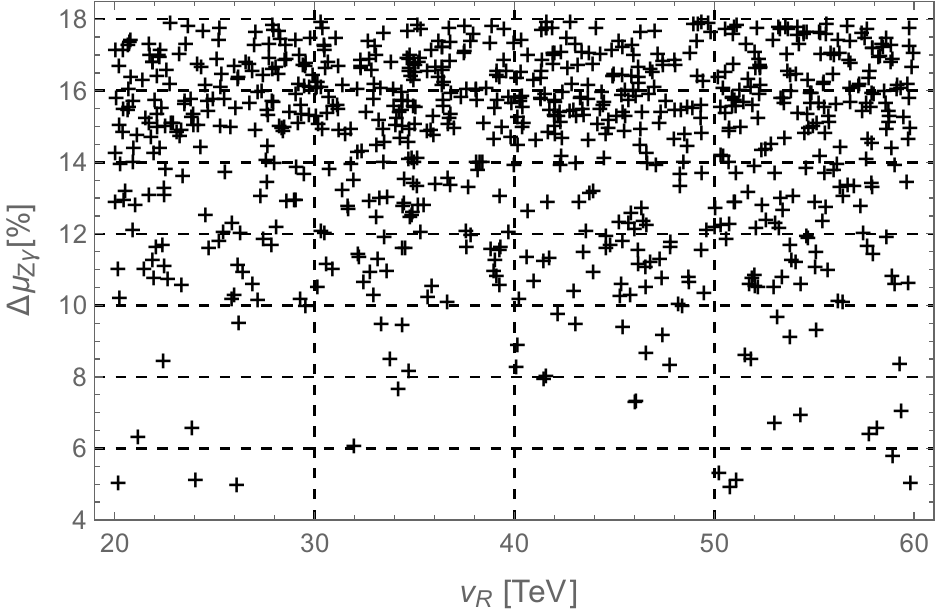}\\ 
	\end{tabular} }
	\vspace*{8pt}
	\caption{ Correlations between $\Delta \mu_{Z\gamma}^{\mathrm{LR}}$ with different values of  $t_{\beta}$ and $v_R$ with $g_R=g_L$. \protect\label{fig_muZga1} }
\end{figure}

For completeness the case of $g_L=g_R$, we discuss on the dependence  of $\Delta \mu^{\mathrm{LR}}_{Z\gamma}$ on $t_{\beta}$ and $v_R$, which  are shown in  Fig.~\ref{fig_muZga1}.
We can see that  $\Delta \mu_{Z\gamma}^{\mathrm{LR}}$  depends weakly on $v_R$, but strongly on $t_{\beta}$. Namely,  all values  $v_R$ can give large $\Delta \mu_{Z\gamma}^{\mathrm{LR}}$, while needs small $t_{\beta}\to 1.2$.

\begin{figure}[ht]
	\centerline{
	\begin{tabular}{cc}
		\includegraphics[width=3.0in]{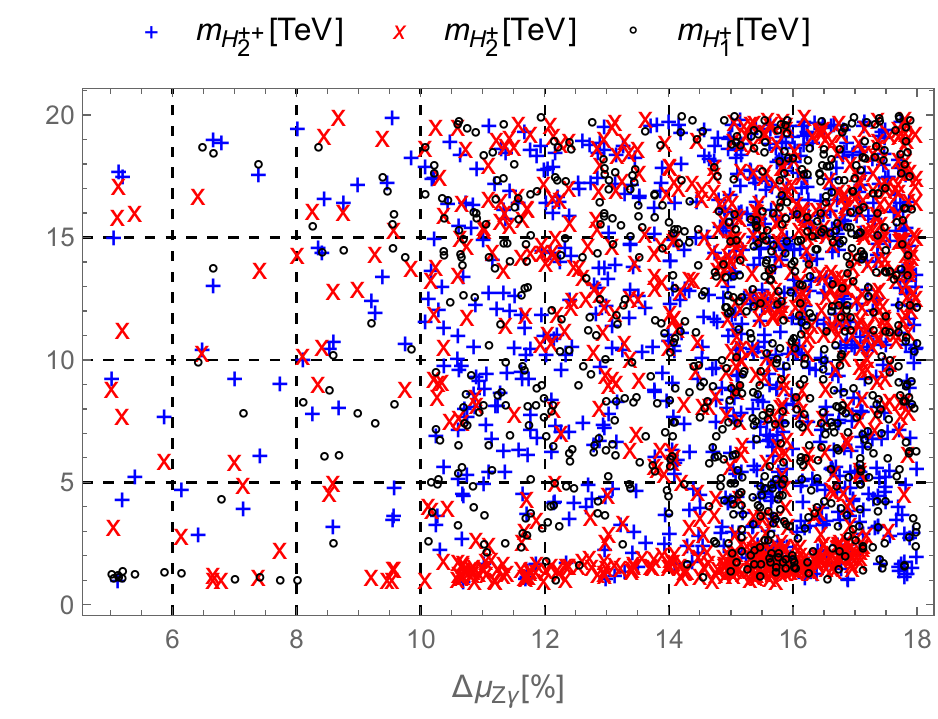}
	\end{tabular} }
	\vspace*{8pt}
	\caption{Correlations between $ \Delta	\mu_{Z\gamma}^{\mathrm{LR}}$ and  charged Higgs bosons masses with $g_R=g_L$.\protect\label{fig_muZga3}}
\end{figure}

The correlations between  $ \Delta \mu_{Z\gamma}^{\mathrm{LR}}$ and  charged Higgs boson masses  are  shown in  Fig. \ref{fig_muZga3}. 
The results  show that all charged Higgs  masses do not affect strongly on values of $\Delta \mu^{\mathrm{LR}}_{Z\gamma}$.

\begin{figure}[ht]
	\centering
	
	\begin{tabular}{cc}
		\includegraphics[width=2.5in]{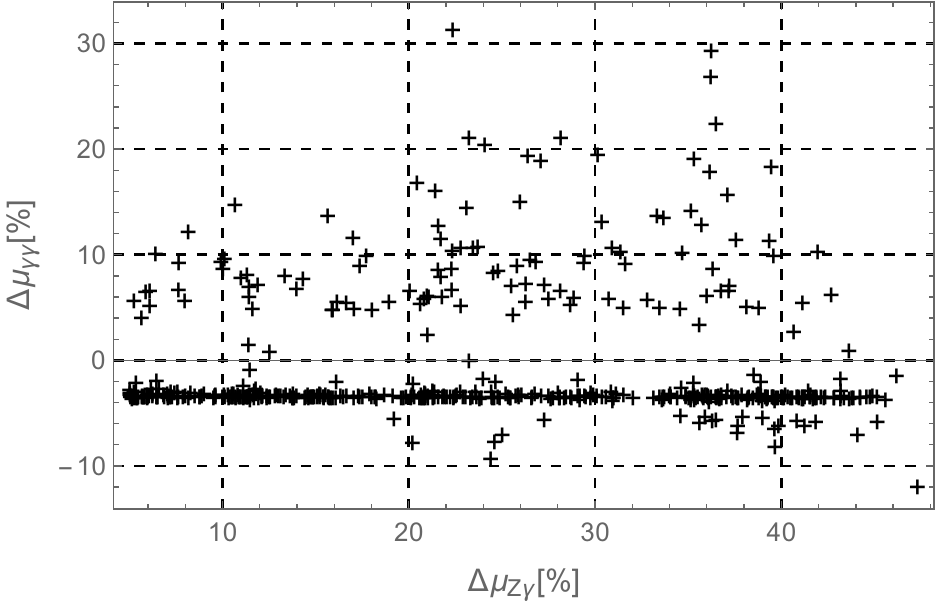}
		\includegraphics[width=2.5in]{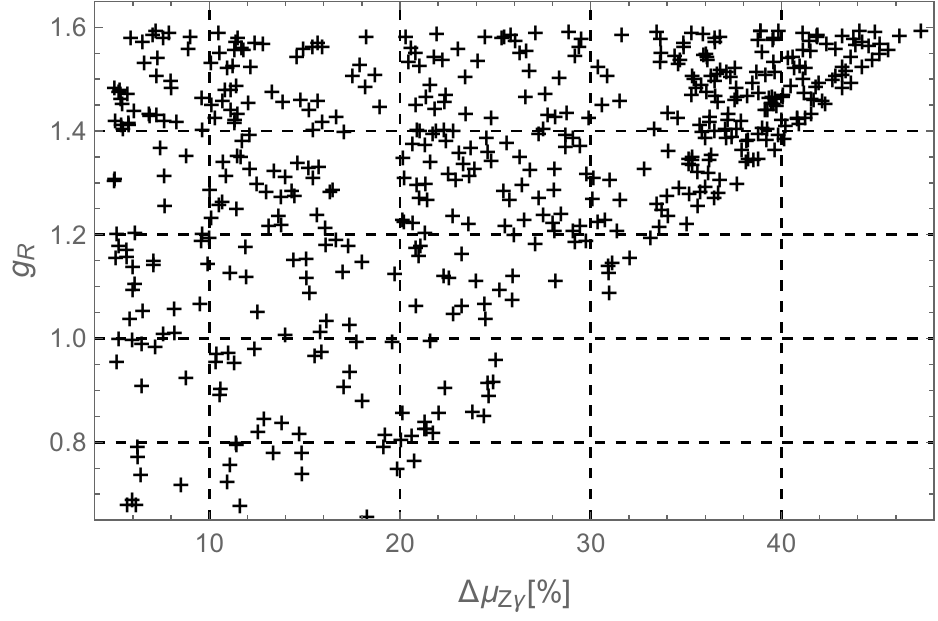}\\ 
	\end{tabular}
	\caption{ Correlations between $\Delta \mu_{Z\gamma}^{\mathrm{LR}}$ with  $\Delta \mu_{\gamma\gamma}^{\mathrm{LR}}$ ($g_R$) in the left (right) panel with $0.65\leq g_R\leq 1.6$. \protect\label{fig_muZgagLR} }
\end{figure}

Finally, we consider the general case of $g_R$ with allowed values given in Eq. \eqref{eq:gRL}.  Numerical results for important correlations between $ \Delta	\mu_{Z\gamma}^{\mathrm{LR}}$ with $ \Delta	\mu_{\gamma \gamma}^{\mathrm{LR}}$ and $g_R$ are depicted in Fig. \ref{fig_muZgagLR}.
It can be seen clearly that  large $\Delta \mu_{Z\gamma}^{\mathrm{LR}}$ corresponds to large $g_R$, which is consistent with the property that new  contributions  consisting of  factor $g_R$ in the Feynman rules  shown in section \ref{sec:coupling}.   We emphasize that lagre $g_R$ is necessary for large $\Delta \mu_{Z\gamma}^{\mathrm{LR}}$ that can reach value of $46\%$, very close to the recent experimental sensitivity.  Furthermore, the expected sensitivity of  $\Delta \mu_{Z\gamma}^{\mathrm{LR}}=4\%$ does not affect  large values of  $\Delta \mu_{Z\gamma}^{\mathrm{LR}}$ that are visible for the incoming experimental sensitivity of    $23\%$. 

\begin{figure}[ht]
	\centering
	
	\begin{tabular}{ccc}
		\includegraphics[width=5.5cm]{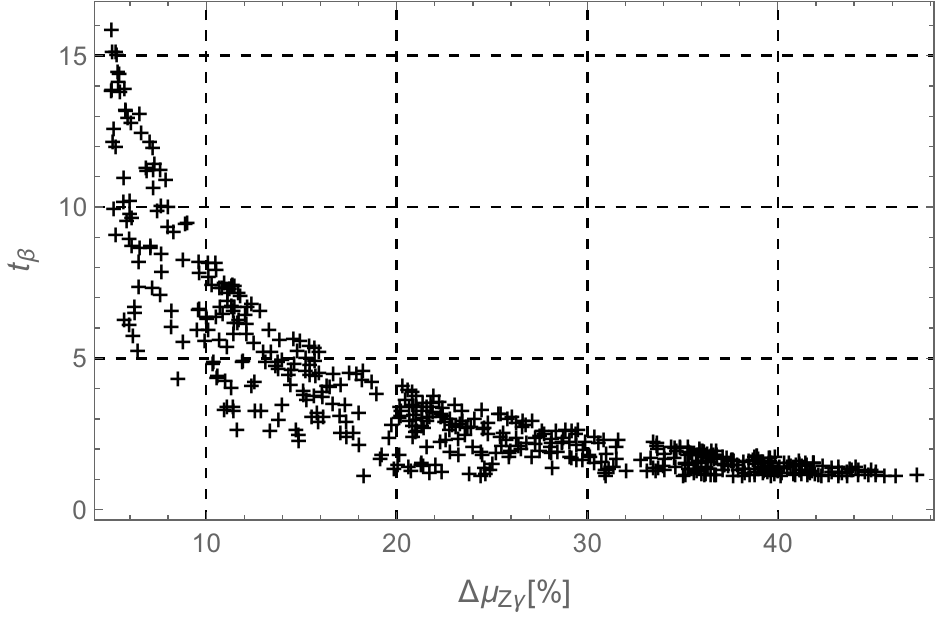}%
			&
		\includegraphics[width=5.5cm]{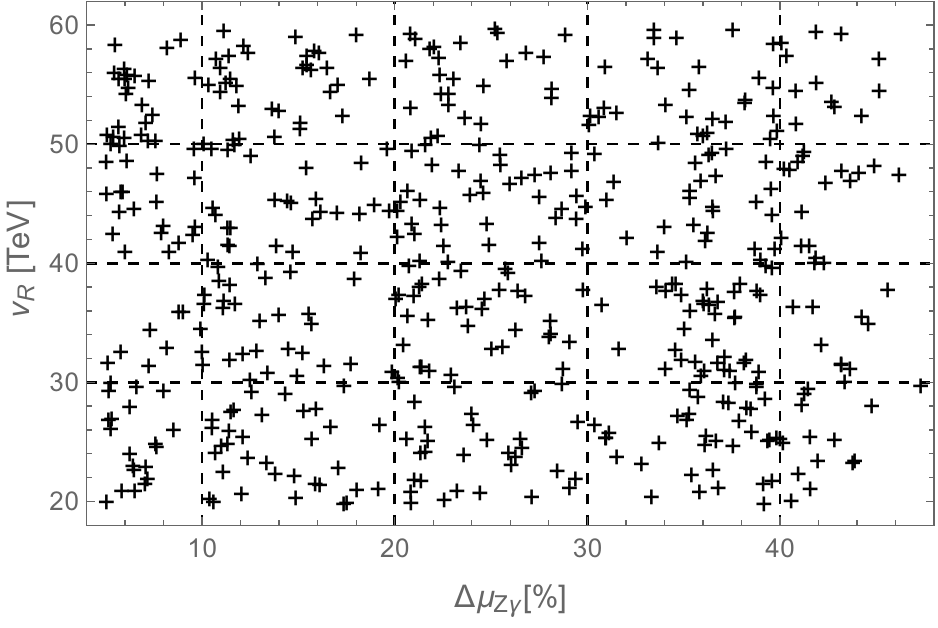}
		& 	\includegraphics[width=5.5cm]{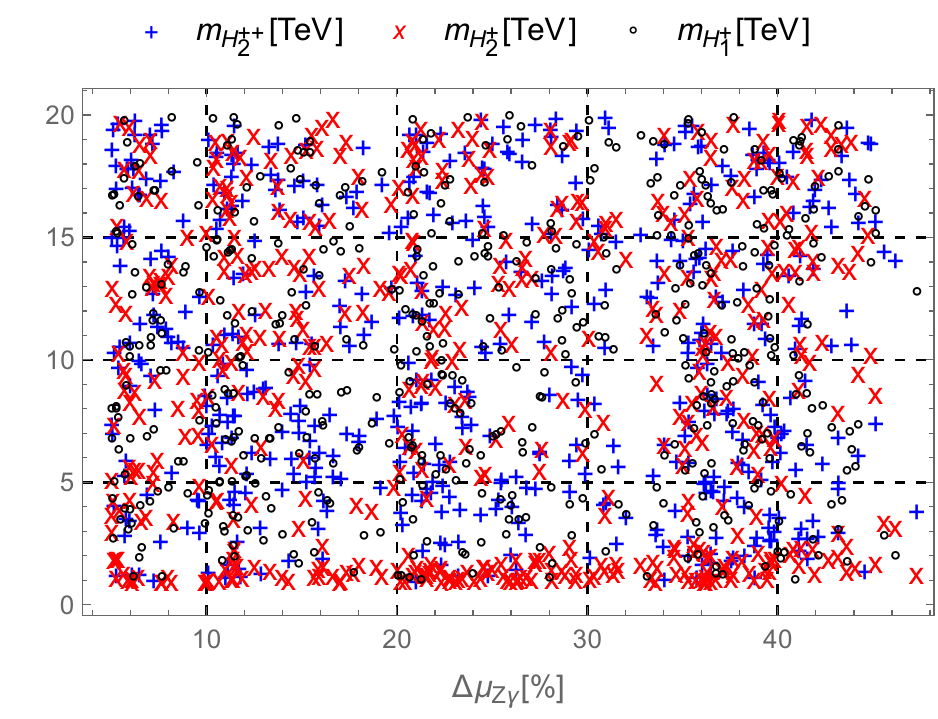}\\ 
	\end{tabular}
	\caption{ Correlations between $\Delta \mu_{Z\gamma}^{\mathrm{LR}}$ with  $t_{\beta}$,  $v_R$ , and charged Higgs masses with $0.65\leq g_R\leq 1.6$. \protect\label{fig_muZgaX1} }
\end{figure}

 Finally, we focus on the correlations between $ \Delta	\mu_{Z\gamma}^{\mathrm{LR}}$ verus  $ t_{\beta}$, $v_R$, and  all charged Higgs  masses, which  are depicted in Fig. \ref{fig_muZgaX1}.
It is seen again that large $ \Delta	\mu_{Z\gamma}^{\mathrm{LR}}$ requires small $t_{\beta}$. In contrast, $ \Delta	\mu_{Z\gamma}^{\mathrm{LR}}$ depends weakly on charged Higgs boson masses and $v_{R}$ given in Eq. \eqref{eq:scanningrange}. 

\section{\label{sec:conclusion} Conclusions}

We have studied all one-loop contributions to  the  SM-like Higgs
 decays $h\to \gamma \gamma ,Z\gamma$ in the MLRSM framework. Interesting properties of the 
 new gauge  and Higgs bosons were explored. Namely,  the SM-like Higgs couplings  were identified with the SM prediction and  experimental data. All masses,  
physical states of  gauge  and Higgs bosons and  their mixing   were presented clearly so that all couplings  relate to one-loop contributions  the decay amplitudes $h \to\gamma\gamma,  Z \gamma$ are derived analytically.  From this, decays $h\rightarrow \gamma\gamma, Z\gamma$ in MLRSM have been discussed using the relevant recent experimental results. The one-loop contributions from the diagrams containing both gauge and Higgs mediation were included in  the decay amplitude $h\rightarrow Z\gamma$.  These contributions were ignored in previous studies, although they may enhance the $h\rightarrow Z\gamma$ amplitude, but do not affect the  $h\rightarrow \gamma\gamma$ one, leading to the possibility that large $\Delta \mu_{Z\gamma}$ may be allowed under the strict experimental constraint of  $\Delta \mu_{\gamma\gamma}$.  We have shown that  the mentioned $h$ decay rates  depend weakly on   $t_\beta$, the $SU(2)_R$ vacuum scale $v_R$. The $2\sigma$ deviation of $\mu_{\gamma\gamma}$ results in a rather strict constraint $\left| \Delta\mu_{Z\gamma}\right|\leq  46\%$. On the other hand,  the large values of $\Delta \mu_{Z\gamma}>23\%$ can appear under  the very strict constraint of $|\Delta \mu_{\gamma\gamma}|\leq 4\%$ corresponding to the future experimental sensitivities,    provided the two requirements that enough  small $t_{\beta}$ and large $g_R$ are necessary. Therefore, the future experimental searches of the two decays mentioned in this work will be important to constrain the parameter space of the MLRSM.

\section*{Acknowledgments}
This research is funded by Vietnam National University HoChiMinh City (VNU-HCM) under grant number “C2022-16-06”.
\appendix

\end{document}